\documentclass[prd,nofootinbib,preprint,showpacs]{revtex4}
\usepackage{graphics}
\usepackage{epsfig}
\usepackage{bm}

\begin{document}

\title{Topology in the Little Higgs Models}

\author{Mark Trodden$^1$\footnote{trodden@physics.syr.edu} and Tanmay 
Vachaspati$^2$\footnote{tanmay@case.edu}}
\affiliation{
$^1$Department of Physics, Syracuse University, 
Syracuse, NY 13244-1130, USA. \\
$^2$CERCA, Physics Department, Case Western Reserve University,
10900 Euclid Avenue, Cleveland, OH 44106-7079, USA.}

\begin{abstract}
We investigate the implications of the nontrivial vacuum structure of little Higgs models.
In particular, focusing on the littlest Higgs model, we demonstrate the existence of
three types of topological defects. One is a global cosmic string that is truly topological. 
The second is more subtle; a semilocal cosmic string, which may be 
stable due to dynamical effects. The final defect is a $Z_2$ monopole solution with an
unusual structure. We briefly discuss the possible cosmological
consequences of such nonperturbative structures, although we note that these 
depend crucially on the fermionic content of the models.
\end{abstract}

\maketitle

\section{Introduction}
\label{sec:intro}

In the last few years, a new partial resolution of the hierarchy problem 
in particle physics has been proposed. This proposal, known as 
{\it the little Higgs} mechanism~\cite{Arkani-Hamed:2001nc,Arkani-Hamed:2001ca,Hill:2000mu,Arkani-Hamed:2002qy,Arkani-Hamed:2002qx,Low:2002ws,Kaplan:2003uc,Chang:2003un,Skiba:2003yf,Chang:2003zn,Katz:2003sn}, introduces new symmetries at the TeV 
scale, preventing the quadratic running of coupling constants and 
postponing the hierarchy problem by at least an order of magnitude in 
energy. Any potential solution to the hierarchy problem deserves serious 
attention and the little Higgs paradigm has therefore generated a lot of
interest.

The basic idea is to make the electroweak Higgs particle a pseudo-Goldstone 
boson of a higher symmetry group. This construction means that quadratic 
divergences in the Higgs mass cancel to one loop. Thus, the Higgs remains 
weakly-coupled beyond the electroweak scale, perhaps up to 10 TeV.

Almost all work on this topic concerns the construction and phenomenological 
analysis of effective theories implementing the little Higgs idea. In this 
letter we adopt a different approach and focus on the nonperturbative structures 
that may be present in specific little Higgs models. Each little Higgs model 
introduces a new symmetry group at the TeV scale. Some of the new symmetries 
are local and some global and the group is spontaneously broken down to that 
of the standard model at electroweak scales. For us this raises the natural 
question of the possible existence of topological defect solutions in the 
low-energy theory.

There exists a wealth of literature concerning topological defects which is 
well summarized in~\cite{vilenkinshellard,Achucarro:1999it}. Here, for completeness, 
we merely provide a lightning review of the main concepts necessary for the 
analysis we perform in the rest of the paper.

Consider a field theory described by a continuous symmetry group $G$ which 
is spontaneously broken to a subgroup $H\subset G$. The space of
all accessible vacua of the theory, the {\it vacuum manifold}, is defined to
be the space of cosets of $H$ in $G$; ${\cal M} \equiv G/H$. 

In general, the theory possesses a topological defect if some homotopy group 
$\pi_i({\cal M})$ is nontrivial, where, in particular, $i=0,1,2,3$ lead to 
domain walls, cosmic strings, monopoles or textures respectively.

Applying these conditions to the electroweak theory, a particularly useful 
example here, the vacuum manifold is the space of cosets
\begin{equation}
{\cal M}_{EW} = \left \{ [SU(2)_L \times U(1)_Y]/Z_2 \right \}/U(1)_Q \ ,
\end{equation}
which is topologically equivalent to the three-sphere, $S^3$, with vanishing 
zeroth, first and second homotopy groups. Thus, the electroweak model does not 
lead to walls, strings, or free magnetic monopoles. 
It does however, lead to texture, which 
exists because $\pi_3({\cal M})$ is nontrivial. It also contains confined magnetic
monopoles \cite{Nambu:1977ag,Achucarro:1999it}.

Further, while the topological structure of the models may be of intrinsic 
particle physics interest, when one considers such models in the context of 
the expanding, cooling universe, the possibility of the cosmic creation of 
topological relics of the little Higgs phase is raised. Such structures may 
serve three possible purposes. They may, of course, have no observable 
consequences in today's universe. Alternatively, they may yield specific 
observable signatures and indeed may even aid in the resolution of some 
cosmological problems. Finally, in the most extreme case, it is possible 
that topological remnants of the little Higgs phase may predict a cosmological 
catastrophe, which may be used to bound specific models. Historically these 
considerations were considered natural in the case of new symmetry groups 
at the grand unified scale and can have interesting effects~\cite{Davis:1997bs}
in the presence
of the other approach to the hierarchy problem - supersymmetry. 
Here we initiate an analogous program of study 
for the little Higgs paradigm.

This paper is organized as follows. In section~\ref{sec:littlest} we consider 
as a prototypical example the {\it littlest Higgs} model and describe its 
symmetry groups. In section~\ref{structure} we examine in detail 
the symmetry breaking structure and in section~\ref{sec:topdef} we 
identify the topological and embedded defects that result.
Finally, in~\ref{sec:conclusions} we summarize our results and comment on the
possible cosmological consequences of the topological aspects of the little
Higgs models.

\section{The littlest Higgs model}
\label{sec:littlest}

We consider the ``littlest Higgs" model \cite{Arkani-Hamed:2002qy}. The model
is constructed by starting with a purely global $SU(5)$ theory for a
scalar field $\Sigma$ that is a $5\times 5$ symmetric, complex matrix.
The transformation law for $\Sigma$ is
\begin{equation}
\Sigma \rightarrow U\Sigma U^T \ ,  \ \ U \in SU(5) \ .
\end{equation}
Note that $U^\dag U = {\bf 1}$ and that $\Sigma$ is multiplied on the
right by $U^T$, not $U^\dag$. The Lagrangian for $\Sigma$ is
\begin{equation}
\label{littlestlag}
{\cal L}={\rm Tr}\left[(\partial_{\mu}\Sigma)^*\partial^{\mu}\Sigma\right]
             -V(\Sigma) \ ,
\end{equation}
where
\begin{equation}
\label{potential}
V(\Sigma)=-\frac{1}{2}\mu^2\Sigma_{ij}^*\Sigma^{ij}+
              \frac{\kappa_1}{4}\left(\Sigma_{ij}^*\Sigma^{ij}\right)^2
+\frac{\kappa_2}{4}\left(\Sigma_{ij}^*\Sigma^{jk}\Sigma_{kl}^*\Sigma^{li}\right) \ ,
\end{equation}
where $\mu$ has dimensions of mass and $\kappa_1$ and $\kappa_2$ are 
dimensionless coupling constants.

The full global symmetry group of this Lagrangian can be seen by noting that 
there is an extra set of transformations, corresponding to multiplication of
$\Sigma$ by $e^{i\alpha}{\bf 1}$, with $\alpha$ a real parameter, under which 
${\cal L}$ is also invariant.
Modding out by the discrete set of symmetries common to both $SU(5)$ and 
to this new $U(1)$ -- namely by the center of $SU(5)$ -- reveals the full 
symmetry group to be
\begin{equation}
\label{Gglobal}
U(5)\cong\frac{[SU(5)\times U(1)]}{Z_5} \ .
\end{equation}

For $5\kappa_1+\kappa_2 > 0$, $\kappa_2>0$, the vacuum expectation value (VEV) 
of $\Sigma$ is $\propto {\bf 1}$ ~\cite{Arkani-Hamed:2002qy,Li:1973mq}. Given the above 
analysis of the gauge groups, the subgroup left unbroken by the above VEV is then 
easily calculated, yielding the full symmetry breaking scheme as
\begin{equation}
\frac{\left[SU(5)\times U(1)\right]}{Z_5} \longrightarrow SO(5)\times Z_2 \ .
\label{su5xu1}
\end{equation}
where the left-over $Z_2$ is due to transformations with ${\rm Det}(U)=\pm 1$.
In other words, the symmetry breaking is $U(5) \rightarrow O(5)$.

The $U(5)$ model above is provided simply as a motivation for the 
construction of the actual model. The next step in constructing the 
littlest Higgs model is to introduce gauge fields. However, instead of
gauging the full $U(5)$ only a $U(2)^2$ subgroup is gauged.
The Lagrangian~(\ref{littlestlag}) becomes
\begin{equation}
\label{littlestgaugelag}
{\cal L}={\rm Tr}\left[(D_{\mu}\Sigma)^*D^{\mu}\Sigma\right]-V(\Sigma) 
          - \frac{1}{4} \sum_{j=1}^2 [ (W_j^{a\mu\nu})^2 + (B_j^{\mu\nu})^2 ] \ ,
\end{equation}
where the covariant derivative is given by
\begin{equation}
\label{covderiv}
D_{\mu}\Sigma =\partial_{\mu}\Sigma - \sum_{j=1}^{2}
\left[ig_j W_{j\mu}^a\left(Q_j^a \Sigma +\Sigma {Q_j^a}^T \right)
+ ig'_j B_{j\mu} \left(Y_j \Sigma+\Sigma Y_j^T \right)\right] \ .
\end{equation}
Here $g_j$ are the coupling constants corresponding to the two $SU(2)$ subgroups 
of $SU(5)$ and $g'_j$ are those corresponding to the two $U(1)$ subgroups, and
$W_j^a$ and $B_j$ are the $SU(2)$ and $U(1)$ gauge fields. Thus 
one has gauged a $[(SU(2)\times U(1))/Z_2]^2 \equiv K$ subgroup of $SU(5)$.  
What remains of the global $U(5)$ symmetry is merely the $U(1)$ factor written
explicitly in Eq.~(\ref{su5xu1}). The generators of the first $SU(2)$, the second 
$SU(2)$, the first $U(1)$ and the second $U(1)$, embedded into $SU(5)$, 
respectively are~\cite{Arkani-Hamed:2002qy},
\begin{eqnarray}
Q_1^a & = & \left(\begin{array}{ccc}
\sigma^a/2 &  0 & 0  \\
0 & 0 & 0 \\
0 & 0 & 0 
\end{array} \right) \ ,\\
Q_2^a & = & \left(\begin{array}{ccc}
0 &  0 & 0  \\
0 & 0 & 0 \\
0 & 0 & -\sigma^{a*}/2  
\end{array} \right) \ ,\\
Y_1 & = &\frac{1}{10}\rm{diag}(-3,-3,2,2,2) \ , \\
\nonumber \\
Y_2 & = &\frac{1}{10}\rm{diag}(-2,-2,-2,3,3) \ ,
\label{generators}
\end{eqnarray}
where $\sigma^a$ are the Pauli spin matrices.

In this way the gauging has diminished the symmetry of the model from $U(5)$
to $K \times U(1)_{g}$ where the subscript $g$ denotes that the
symmetry is global. Note that the potential of the model continues to be
given by Eq.~(\ref{potential}) and still carries the full $U(5)$ 
symmetry. This fact is important for us and we will describe its consequences
more fully in the next section.

\section{Vacuum structure of the model}
\label{structure}
The Lagrangian~(\ref{littlestgaugelag}) of the little Higgs model in has 
the peculiar feature that the symmetries of the gradient and potential terms
are different. Before proceeding to elaborate this feature, let us consider
the standard electroweak model where a similar feature is present. 

In the electroweak model, the potential is
\begin{equation}
V_{ew}(\Phi ) = \lambda (\Phi^\dag \Phi - \eta^2 )^2 \ ,
\end{equation}
where $\Phi^T = (\phi_1+i\phi_2, \phi_3+i\phi_4)$ is the standard model
Higgs field and $\phi_i$ are its real-valued components. Therefore we can write
\begin{equation}
V_{ew}(\Phi ) = \lambda \left (\sum_{i=1}^{4}\phi_i^2  - \eta^2 \right )^2 \ ,
\end{equation}
from which it is clear that the continuous symmetry of the potential is $SO(4)$,
which is isomorphic to $SU(2)_L\times SU(2)_R$, with subscripts
conforming to conventional notation\footnote{$SU(2)_R$ is
often called the ``custodial'' symmetry.}. Of course, in the electroweak theory 
only the $SU(2)_L$ factor and a $U(1)$ subgroup of the second $SU(2)_R$
factor are gauged. This diminishes the symmetry of the model to the electroweak
symmetry group $[SU(2)_L\times U(1)_Y]/Z_2$. However, since the gauging 
affects only the gradient terms, the potential retains the full $SO(4)$ symmetry. 
Once $\Phi$ gets a VEV, 
this symmetry is spontaneously broken down to $SU(2)_{L+R}$, while the 
gauged electroweak symmetry group is broken down to $U(1)_Q$. These 
considerations may be summarized in the following chart
\begin{equation}
\begin{array}{cccc}
{\rm Global:} & [SU(2)_L \times SU(2)_R]/Z_2 & \rightarrow & SU(2)_{L+R}      
\cr
 \downarrow     & \downarrow       & &    \downarrow   \cr
{\rm Gauge:} & [SU(2)_L \times U(1)_Y]/Z_2  & \rightarrow  &  U(1)_Q           
\end{array}
\end{equation}

\begin{figure}[ht]
\centering
\includegraphics[width=2.5in]{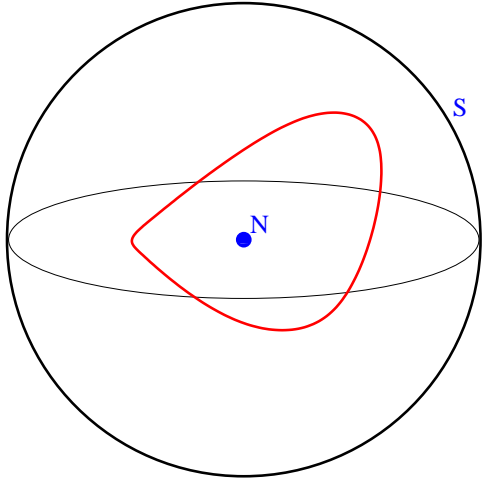}
\caption{The vacuum manifold of the standard electroweak model is an $S^3$.
If only the $U(1)$ group of the electroweak symmetry group is gauged, the
manifold is better viewed as the Hopf fibration of $S^3$ \cite{Gibbons:1992gt}.
Then there are topologically non-trivial paths corresponding to orbits of the 
gauged $U(1)$. Field configurations that lie in the gauge orbit are trivial 
in the full $S^3$, yet to deform them to a trivial field configuration requires 
gradients that cannot be compensated by the gauge field. Hence the deformation
costs energy and this is why topological defects corresponding to the
gauged orbits can be stable even in the full model. Such defects are
called ``semilocal''.}
\label{ewfigure}
\end{figure}

Now consider the vacuum structure of the electroweak model. $V_{ew}(\Phi )$
is minimized on a three sphere, namely, $\sum \phi_i^2 = \eta^2$. The manifold
obtained by symmetry considerations, $\{[SU(2)_L\times U(1)_Y]/Z_2\}/U(1)_Q$
is also three dimensional but it is important to realize that this is merely 
a coincidence. We shall see that in the littlest Higgs model, in which a 
similar structure exists, such a coincidence does not occur. The 
electroweak model does not contain any topological defects. However, 
if $SU(2)_L$ is left ungauged the theory does contain stable semilocal 
defects~\cite{Vachaspati:dz,Hindmarsh:1991jq,Preskill:1992bf,Achucarro:1999it} 
for the following reason. As above, the minimum of the potential remains an 
$S^3$, on which there are special orbits that are gauged (Fig.~\ref{ewfigure}).
(If there are gradients of $\Phi$ along these orbits, they can be completely 
compensated by a gauge field.) These orbits are $S^1$'s and hence are 
topologically non-trivial. A field configuration that lies on such a 
gauge orbit can be deformed to a constant field everywhere but the deformation 
costs gradient energy since the configuration needs to be lifted off the 
orbit and contracted on the $S^3$. If the scalar couplings are large compared 
to the gauge couplings, then this energy cost is large enough to stabilize
the field configuration, as has been explicitly seen in the
case of semilocal and electroweak strings. Since the defect is stable
due to an interplay between global and gauge symmetries, such defects are 
called semilocal cosmic strings. For the electroweak model with $SU(2)_L$ 
left ungauged, the electroweak semilocal string is stable provided
$m_H^2 < m_Z^2$ where $m_H$ and $m_Z$ are the masses of the Higgs scalar and the
vector boson.

We now move on to the littlest Higgs model. Here the potential $V(\Sigma )$ is 
invariant under the full $U(5)$ group, while the Lagrangian as a whole respects
a $K \times U(1)_{g}$ symmetry. 

\begin{figure}[ht]
\centering
\includegraphics[width=2.5in]{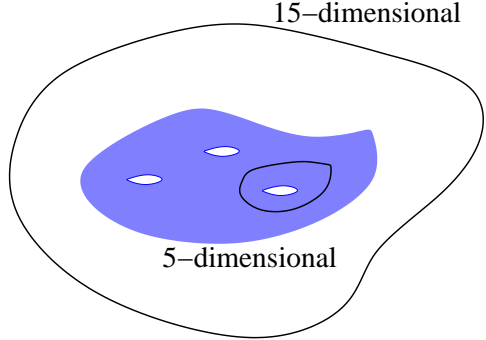}
\caption{The minimum of the potential in the littlest Higgs model is a 15
dimensional space and the gauge orbits form a 5 dimensional subspace. The
5 dimensional subspace may have non-trivial topology but the corresponding
semilocal defects will be stable only if the gradient cost in lifting the
field configurations off the gauge orbit is significant.}
\label{semilocalfigure}
\end{figure}

The potential is minimized on a $25-10=15$ dimensional manifold 
$M_V\equiv U(5)/O(5)$. 
However, the vacuum manifold $M_L$ corresponding to the full Lagrangian
is isomorphic to a coset space 
\begin{equation}
M_L \cong \frac{K\times U(1)_{g}}{U(2)\times Z_2} \ . 
\end{equation}
This is only a $9-4=5$ dimensional space. As depicted
in Fig.~\ref{semilocalfigure}, $M_V \supset M_L$
and so non-trivial topological features of $M_L$ need not 
be non-trivial in $M_V$. Therefore, in order to discuss topological defects in the 
model we need
to (i) find any non-trivial topological features of $M_L$ and (ii) check if
this non-trivial topology in $M_L$ can be trivialized in the full manifold $M_V$. 
Of course, if such a trivialization is not 
possible then the topological defect solution
in $M_L$ is a genuine topological defect in the full theory. However, if the
topology of the configuration is trivial in $M_V$, then the topological defect in $M_L$
may still be a semilocal defect of the full theory and may be stable over 
some parameter range.
As we shall now see, the littlest Higgs model has both topological and semilocal
strings.

The chart for this model corresponding 
to the one shown above for the electroweak model is:
\begin{equation}
\begin{array}{cccc}
{\rm Global:}  & [SU(5)\times U(1)_{g}]/Z_5 & \longrightarrow & SO(5) \times Z_2    \cr
\downarrow                           & \downarrow     &  & \downarrow  \cr
{\rm Gauge:}   & K \times U(1)_{g} & \longrightarrow & 
                     \{ [SU(2)_L\times U(1)_Y ]/Z_2 \} \times Z_2   \\
\end{array}
\label{su5chart}
\end{equation}
where, as defined earlier,
\begin{equation}
K =\{[SU(2)\times U(1)]/Z_2 \}^2 \ .
\end{equation}

\section{Topological defects in the littlest Higgs model}
\label{sec:topdef}
For the Lagrangian~(\ref{littlestgaugelag}) the structure of the unbroken symmetry 
groups was best elucidated in a basis in which 
$\langle\Sigma\rangle\propto {\bf 1}$ (see the discussion below Eq.~(\ref{Gglobal})). 
However, at this stage it is simpler to use a different basis in which the VEV is given by
\begin{equation}
\label{VEV}
\langle\Sigma\rangle=\Sigma_0\equiv  \left(\begin{array}{ccc}
0_{2\times 2} &  0_2 & 1_{2\times 2}  \\
0_2 & 1 & 0_2 \\
1_{2\times 2} & 0_2 & 0_{2\times 2} 
\end{array} \right) \ .
\end{equation}
The symmetry breaking scheme is still as shown in Eq.~(\ref{su5chart}) and it 
is straightforward to check that the unbroken generators corresponding to the 
$SU(2)_L$ remaining after the symmetry breaking are $Q_1^a+Q_2^a$ 
(see Eq.~(\ref{generators})).
Similarly the unbroken $U(1)_Y$ has generator $Y_1+Y_2$. 

\subsection{String Solutions}
We are now in a position to find the topological defects in the littlest Higgs model. First 
consider the spontaneous breaking of the $U(1)_{g}$ factor in the symmetry breaking. 
The $U(1)_{g}$ factor spontaneously breaks to a $Z_2$ subgroup. Since $\pi_1(U(1)/Z_2) = Z$,
the group of integers under addition, it is clear that the model contains global $U(1)$ 
strings. An appropriate ansatz for these objects is given by
\begin{equation}
\label{globalstring}
\Sigma_{g}(r,\theta) =f_{g}(r)e^{i\theta}\Sigma_0 \ ,
\end{equation}
where $f_{g}(r)$ is a real profile function for the global string.

Similarly, in the gauged $U(1)$ sector there is a cosmic string configuration corresponding 
to the broken $U(1)$ symmetry whose generator is $Y_1-Y_2$ which is orthogonal to $Y_1+Y_2$,
the generator of $U(1)_Y$. Let us define
\begin{equation}
{\tilde Y} = 10 (Y_1 - Y_2) ={\rm diag}(-1,-1,4,-1,-1) \ .
\end{equation} 
Then an ansatz for the string is
\begin{equation}
\label{localansatz}
\Sigma_l(r,\theta)=f_l(r)e^{i{\tilde Y} \theta }\Sigma_0 \ , \ \ 
A_\theta = \frac{v_l(r)}{g'r} \ ,
\end{equation}
where, in terms of the hypercharge gauge fields in Eq.~(\ref{covderiv}),
\begin{equation}
A_\theta = \cos(\alpha ) B_{2\theta} - \sin(\alpha) B_{1\theta}
\end{equation}
and
\begin{equation}
g ' \equiv\sqrt{{g'}_1^2+{g'}_2^2} \ , \ \ \tan (\alpha)\equiv \frac{g_1'}{g_2'} \ ,
\end{equation}
with all other fields vanishing. Here $f_l(r)$ and $v_l(r)$ are real profile functions, 
obeying the usual second order ordinary differential equations satisfied by the 
Nielsen-Olesen vortex, with boundary conditions 
\begin{equation}
f_l(0)=\frac{dv_l}{dr}(0)=v_l(\infty)=0  \ , \ \ 
f_l(\infty)=  \left[\frac{1}{5\kappa_1 +\kappa_2}\right]^{1/2}\mu \ .
\end{equation}

This string has a tension $T\sim (10~{\rm TeV})^2$ and carries a magnetic flux 
associated with the gauge field $A_{\mu}$ given by
\begin{equation}
\Phi_A =-\frac{2\pi}{g'} \ .
\end{equation}

Note that these solutions may not be the least energy solutions
for the given topology. For example, the presence of bosonic condensates could change 
the energy per unit length. However, the existence of a condensate is a model-dependent
question.

This second string is not apparent in the global symmetry breaking shown in the
top line of Eq.~(\ref{su5chart}) since $\pi_1 (SU(5)/SO(5))$ is trivial. Clearly this
is because the
string configuration can be deformed to the trivial configuration by using
transformations belonging to the full $SU(5)$. Thus the second string
is a semilocal string.

The stability of the semilocal string depends on the relative importance of 
the potential energy to the gradient energy. If the scalar coupling constants are
large, it is favorable to minimize the potential energy even at some cost of
gradient energy. If the gauge coupling constant is large, it is most favorable
to have vanishing gradient energy. 

The stability condition that is analogous to the electroweak condition is that
$m_s^2 < m_v^2$ where $m_s$ and $m_v$ are the scalar and vector masses in 
the model. In terms of the coupling constants:
\begin{equation}
\frac{5}{2}\left(\frac{5\kappa_1 +\kappa_2}{{g'}_1^2+{g'}_2^2}\right)<1 \ .
\end{equation}

\subsection{A Monopole Solution}

We now turn briefly to another defect solution in the littlest Higgs model. 
Although the symmetry breaking scheme is somewhat complicated, 
as described in~(\ref{su5chart}), it is clear that a subset of the 
scheme involves $SU(5)\rightarrow SO(5)$, effected by the symmetric 
tensor representation $\Sigma(x)$. However, it is relatively simple to show 
that
\begin{equation}
\label{pi2eqn}
\pi_2\left[SU(5)/SO(5)\right] = Z_2 \ ,
\end{equation}
(for example by using the exact homotopy sequence which yields 
$\pi_2[SU(5)/SO(5)] =\pi_1[SO(5)]$).
Hence, the theory contains a particularly interesting type of magnetic 
monopole -- a $Z_2$ monopole (see for example~\cite{Preskill:1986kp}). 
Because these monopoles correspond to a $Z_2$ homotopy group, the 
monopole and the antimonopole are identical.

If the only symmetry breaking was the global $SU(5)\rightarrow SO(5)$, 
then these monopoles would carry a purely global non-Abelian $SO(5)$ 
charge. However, since the littlest Higgs model involves gauging subgroups 
of the $SU(5)$, the resulting monopoles will be partially gauged.
Consider an $SU(3)$ subgroup of $SU(5)$ that lies in the upper
$3\times 3$ block of $SU(5)$ group elements.
When $\Sigma$ gets a VEV proportional to ${\bf 1}$, the
$SU(3)$ subgroup breaks down to $SO(3)$, and this symmetry breaking
too leads to $Z_2$ monopoles. Therefore, to construct the monopoles,
we need only consider the symmetry breaking $SU(3) \rightarrow SO(3)$.
Now an $SU(2)\times U(1)$ subgroup of this $SU(3)$ is gauged. But
this is the maximal non-trivial subgroup of $SU(3)$. The 
scalar field of a $Z_2$ monopole on the two sphere at infinity, is 
given by the action of elements of $SU(3)$ on some chosen VEV, say
proportional to ${\bf 1}$. However, since none of the $SU(3)$
elements commutes with all the elements of the
$SU(2)\times U(1)$ maximal subgroup, at least some of the angular 
gradients can be compensated by suitable gauge fields. This is
what we mean by ``partial gauging''. An explicit solution of the
partially gauged $SU(3) \rightarrow SO(3)$ monopole is not known
but would be very interesting to work out, especially if
the monopole carries some electromagnetic field distribution 
after the electroweak symmetry breaking.

\section{Conclusions and Comments}
\label{sec:conclusions}
Little Higgs models provide a new logical possibility for explaining the 
stability of the weak scale. In such models the hierarchy problem is postponed 
by an order of magnitude, pushing the relevant scale up to around $10$ TeV. 
In this paper we have investigated the nonperturbative structure of these models, 
in particular their topological defect structure, as a complement to detailed studies 
of their low-energy phenomenology. This seems to be a natural study to perform, 
since little Higgs models make use of new gauge and global symmetries at the 
TeV scale, and thus inherently involve new symmetry breaking schemes that one 
expects to be realized during the thermal evolution of the expanding universe.

Given the large number of possible ways to implement the little Higgs paradigm, 
we have chosen to focus on one of the simplest such examples, the {\it littlest Higgs} 
model. Our analysis demonstrates the existence of three distinct structures. The 
first is a global abelian cosmic string solution, topologically stable, arising 
from the breakdown of a global $U(1)$ symmetry. The second is a more subtle object;
a semilocal gauge string, embedded in the larger group structure of the gauge 
sector of the theory.  This object is not topologically stable, but may be stable 
dynamically, depending on the values of the coupling constants in the theory.
The final object is what we describe as a partially gauged $Z_2$ monopole, which
is also topologically stable.

We have constructed the appropriate ans{\" a}tze for the scalar fields and the 
gauge fields making up these defects. In the case of the semilocal string we 
have also identified the gauge flux carried by the defect.

What remains to be done is a careful analysis of the possible cosmological 
implications of our findings.  
Clearly TeV-scale strings  and monopoles will have negligible gravitational effects 
(for example, one does not expect them to play a role in structure formation.) 
However, the microphysics of such objects may be important in some circumstances. One
example is their potential role in weak scale baryogenesis~\cite{Brandenberger:1988as,Brandenberger:1991dr,Brandenberger:ys,Brandenberger:1994bx}. Another interesting possibility arises if the strings
are superconducting. As originally pointed out by Witten~\cite{Witten:eb}, it is possible for 
cosmic strings to carry supercurrents along them. These may be due to the presence of a scalar
condensate on the string, to fermion zero modes along it, or even more exotic types of superconductivity.
The evolution of a network of such superconducting cosmic strings can differ from a nonsuperconducting one. In particular, the supercurrent along loops of string builds up as the loop radiates away its energy.
This can affect the endpoint of loop evolution. In some cases the supercurrent can become large 
enough to destabilize the loop. In others, the current can compete with the tension of the string loop
and result in stable remnants, known as {\it vortons}~\cite{Davis:ij} that 
constrain the theory~\cite{Brandenberger:1996zp,Carter:1999an} or even may act 
as dark matter~\cite{Brandenberger:1996zp,Martins:jg}. These 
latter suggestions depend crucially on the fermionic content of the theory, and 
particularly on the 
potential existence of fermion zero modes on the strings, leading to superconductivity.
Such effects require a model-specific analysis that is beyond the scope of this 
paper and which we therefore reserve for future work.

Although we have also demonstrated the existence of a partially gauged 
$SU(3) \rightarrow SO(3)$ monopole, as we have commented, the explicit
solution is not known. An interesting future direction is to explicitly 
construct such a solution, since it is possible that the monopole carries 
some electromagnetic field distribution after the electroweak symmetry 
breaking, perhaps resulting in cosmological consequences.

\acknowledgments
MT thanks Graham Kribs, John Terning and Scott Thomas for encouragement to write 
this work up. We also thank John Terning for helpful comments on a first draft of the 
manuscript and Ruth Gregory and Laura Mersini for comments in the early stages of 
the project. The work of MT is supported in part by the National Science Foundation (NSF) 
under grant PHY-0094122. MT is a Cottrell Scholar of Research Corporation. 
TV's work was supported by the DOE.

\end{document}